\documentclass[unnumsec,webpdf,contemporary,large]{oup-authoring-template}%
\graphicspath{{}}

\theoremstyle{thmstyleone}%
\theoremstyle{thmstyletwo}%
\theoremstyle{thmstylethree}%

\begin{document}

\journaltitle{This is a preprint version, and the manuscript has been Submitted to journal}
\DOI{xxxxx}
\copyrightyear{2023}
\pubyear{2023}
\access{Advance Access Publication Date: 15 May 2023}
\appnotes{Research Article}

\firstpage{1}

%\subtitle{Subject Section}

\title[StoManager1: Automating Stomata Metrics]{StoManager1: An Enhanced, Automated, and High-throughput Tool to Measure Leaf Stomata and Guard Cell Metrics Using Empirical and Theoretical Algorithms}

\author[a,$\ast$]{Jiaxin Wang\ORCID{0000-0003-4808-5085}}
\author[a]{Heidi J. Renninger\ORCID{0000-0002-2485-9835}}
\author[b]{Qin Ma\ORCID{0000-0002-6995-6663}}
    \author[c]{Shichao Jin\ORCID{0000-0003-1150-336X}}

\authormark{Jiaxin Wang et al.}

    \address[a]{\orgdiv{Department of Forestry}, \orgname{Mississippi State University}, \orgaddress{\street{Mississippi State}, \postcode{39762}, \state{MS}, \country{USA}}}
\address[b]{\orgdiv{School of Geography}, \orgname{Nanjing Normal University}, \orgaddress{\street{Nanjing}, \postcode{210023}, \state{Jiangsu}, \country{China}}}
\address[c]{\orgdiv{Plant Phenomics Research Centre, Academy for Advanced Interdisciplinary Studies, Collaborative Innovation Centre for Modern Crop Production co-sponsored by Province and Ministry}, \orgname{Nanjing Agricultural University}, \orgaddress{\street{Nanjing}, \postcode{210095}, \state{Jiangsu}, \country{China}}}

\corresp[$\ast$]{Corresponding author. \href{email:jw3994@msstate.edu}{jw3994@msstate.edu}}

\received{15}{5}{2023}
\revised{15}{5}{2023}
\accepted{15}{5}{2023}

%\editor{Associate Editor: Name}

%\abstract{
%\textbf{Motivation:} .\\
%\textbf{Results:} .\\
%\textbf{Availability:} .\\
%\textbf{Contact:} \href{name@email.com}{name@email.com}\\
%\textbf{Supplementary information:} Supplementary data are available at \textit{Journal Name}
%online.}

\abstract{Automated stomata detection and measuring are vital for understanding plant physiological performance and ecological functioning in global water and carbon cycles. Current methods are laborious, time-consuming, prone to bias, and limited in scale. We developed StoManager1, a high-throughput tool utilizing empirical and theoretical algorithms and convolutional neural networks to automatically detect, count, and measure over 30 stomatal and guard cell metrics, including stomata and guard cell area, length, width, and orientation, stomatal evenness, divergence, and aggregation index. These metrics, combined with leaf functional traits, explained 78\% and 93\% of productivity and intrinsic water use efficiency (iWUE) variances in hardwoods, making them significant factors in leaf physiology and tree growth. StoManager1 demonstrates exceptional precision and recall (mAP@0.5 over 0.993), effectively capturing diverse stomatal properties across various species. StoManager1facilitates the automation of measuring leaf stomata, enabling broader exploration of stomatal control in plant growth and adaptation to environmental stress and climate change. This has implications for global gross primary productivity (GPP) modeling and estimation, as integrating stomatal metrics can enhance comprehension and predictions of plant growth and resource usage worldwide. StoManager1's source code and an online demonstration are available on GitHub (https://github.com/JiaxinWang123/StoManager.git), along with a user-friendly Windows application on Zenodo (https://doi.org/10.5281/zenodo.7686022).}

% \boxedtext{
% \begin{itemize}
% \item Key boxed text here.
% \item Key boxed text here.
% \item Key boxed text here.
% \end{itemize}}

\maketitle

\section{Introduction}
Plants regulate hydrological and energy balance by opening and closing stomata for water transpiration and $\mathrm{CO}_2$ uptake, and this process enables plants to complete photosynthesis, maintain water status and leaf temperature, and adapt to climate change (Farquhar \& Sharkey, 1982; Hetherington \& Woodward, 2003; Matthews \& Lawson, 2018). Stomata account for 95\% of terrestrial gaseous fluxes of water and carbon, and studies suggest that stomatal closure induced by elevated atmospheric $\mathrm{CO}_2$ concentration contributes to about 20\% of the global precipitation reduction (Field et al., 1995; Betts et al., 2004; Matthews \& Lawson, 2018). In terms of carbon, about one-sixth of the atmospheric carbon passes through leaf stomata to be fixed annually by terrestrial photosynthesis (Ciais et al., 1997). With changing climatic conditions, there is a growing need to quantify the control of stomata in photosynthetic $\mathrm{CO}_2$ fixation and plant water relations. Understanding stomatal function is crucial in evaluating how plants react to environmental stress, particularly when faced with limited water supply, and it is essential in identifying plants that have reduced water usage yet can achieve high yields in harsher environments (Morison et al., 2008; Lawson, 2009). To these ends, more efforts on studying leaf stomata are warranted.

Stomata are morphologically and mechanically diverse across different species, and variation of stomatal characteristics enables plant's diverse functionalities and adaptation to environmental conditions (Hosy et al., 2003; SchluÈter et al., 2003; Franks \& Farquhar, 2006; Pearce et al., 2006; Orsini et al., 2012; Chen et al., 2017). Commonly used morphological indicators are stomatal density and stomatal size or length. Stomatal density represents the number of stomata per unit leaf surface area, which can be affected by leaf development and environmental factors (Beerling \& Chaloner, 1993; Brownlee, 2001). For example, Woodward and Kelly (1995) observed a 14.3\% stomatal density reduction with $\mathrm{CO}_2$ enrichment. Woodward (1987) found a 40\% decrease in the stomatal density of leaves from eight temperate arboreal species collected over the last 200 years. Variations in stomatal density can also affect a plant's physiological performance. For instance, Franks et al. (2015) improved the water-use efficiency of $Arabidopsis$ $thaliana$ (L.) by genetically manipulating stomatal density. Plants can also change their stomatal density to adapt to challenging environmental conditions. Hughes et al. (2017) reported that barley improves its drought tolerance by reducing leaf stomatal density without affecting yield. Stomatal size affects their response to environmental signals, and small stomata are generally assumed to respond more quickly (Elliott-Kingston et al., 2016). Stomatal density and size are measured manually or through semi-automatic labeling and segmentation by photographing images under a microscope and counting and measuring stomata using image software (Omasa \& Onoe, 1984; Minervini et al., 2015; Rueden et al., 2017). However, these methods are laborious, time-consuming, and limited to small datasets as well as image size and quality, which may cause bias and ambiguities. These factors affect our understanding of stomatal characteristics and functioning and prevent their use in large-scale studies and process-based models.

An easy-to-use methodology for high-throughput and automatic measurement of stomatal characteristics, such as stomatal density, size, area, and orientation, is needed to fully explore stomatal parameters regarding water regulation and carbon cycling. To this end, computer vision technology can be a potentially powerful tool. Recently, studies have developed some methods for stomatal detection and counting using state-of-the-art machine learning algorithms, such as convolutional neural network (CNN) (Toda et al., 2018; Fetter et al., 2019; Casado-García et al., 2020; Liang et al., 2022; Ott \& Lautenschlager, 2022). Although these methods proved the possibilities and feasibility of using machine learning for stomatal detection and counting, none of them perform comprehensive measurement of stomatal metrics. Specifically, it is unclear if stomatal metrics, especially the metrics that are laborious to measure, such guard cell width, length, and area, could be captured automatically, precisely, and sensitively. In addition, the quality of the image used to analyze stomata can impact the accuracy of the measurements. If the image is blurry or has low resolution, it is difficult to accurately identify and measure individual stomatal characteristics (Sultana et al., 2021). Furthermore, stomatal morphological features can be influenced by various factors, such as genetic differences between individuals or environmental conditions, which may cause variation and make the accurate and precise measurement of stomatal characteristics challenging (Hong et al., 2018). As a result, accurately capturing the variation in stomatal characteristics can be challenging. Additionally, some of the currently available methods for measuring stomatal characteristics are hindered by technical difficulties, such as the requirement of programming skills, which can be a barrier for researchers outside of the field (Ott \& Lautenschlager, 2022). Therefore, it is essential to continue to develop and refine measurement methods for stomatal characteristics to improve their accuracy and sensitivity. Since the research objectives and objects can vary across different scientific communities, building an easily used methodology that can be implemented broadly, or at least is not species specific, is crucial.

CNNs have been extensively studied among deep neural networks, which have heavily accelerated computer vision development and implications (Gu et al., 2018). CNNs are primarily used to resolve challenging, image-driven pattern recognition tasks. With precise, yet straightforward architecture, CNNs present an easy entry point to artificial neural networks (ANNs) for computer vision (O'Shea \& Nash, 2015). ANNs are computational processing systems inspired by biological nervous systems, with many interconnected computational nodes that connect to learn from the input and optimize the final output (O'Shea \& Nash, 2015). Redmon et al. (2016) proposed 'You Only Look Once (YOLO)' and published an improved version of YOLOv3, which has been broadly accepted and applied in computer vision for its fast speed in training, detection, and accuracy (Redmon \& Farhadi, 2017; Redmon \& Farhadi, 2018; Du et al., 2020; Liu \& Wang, 2020; Zhang et al., 2022). Instead of using region proposal methods to generate bounding boxes in an image to classify and refine the bounding boxes for object detection and scoring, YOLO reframes object detection as a single regression problem by straightening from image pixels to bounding box coordinates and class probabilities. This process makes YOLO 100-1000 times faster than other approaches like region-based convolutional neural networks (R-CNN). Hence, it is an excellent opportunity to adopt this state-of-the-art, deep-learning algorithm for high-throughput stomatal characterization by building accessible and ready-to-use applications.

In this study, we employed microscope images from two distinct datasets - the hardwood dataset, which encompasses 16 hardwood species, and the $Populus$ dataset, which includes seven taxa and 55 genotypes - to develop and assess YOLO models (bounding box- and segmentation-based models) (Table 1). We hypothesized that YOLO models could be leveraged to detect, count, and measure stomata. Additionally, based on the output, we can develop new metrics that can be used to explain leaf physiological and growth performance. Specifically, we aimed to (1) train versatile models capable of detecting and/or segmenting leaf stomatal pore and the stomatal pore and guard cell complex in both $Populus$ and other hardwood species, (2) utilize empirical and theoretical algorithms to estimate and/or measure stomatal metrics such as stomatal density, guard cell width, length, area, the ratio of stomata pore area/guard cell area, and (3) explore the biological application of developed stomatal metrics, such as stomatal orientation, guard cell area, the ratio of stomata pore area/guard cell area, and the variance in stomatal area, orientation, guard cell width, length, and the ratio of stomata pore area/guard cell area. With our framework, plant scientists and ecologists can efficiently conduct large-scale and comprehensive studies of leaf stomata enabling greater investigation into the characteristics of stomata and their roles in regulating plants' carbon gain and water loss more efficiently.

\section{Materials and Methods}\label{sec2}

\subsection{Leaves and micrographs}\label{subsec1}
This study utilized stomatal images from the hardwood and $Populus$ spp. data sets, which were collected and prepared in 2015 and 2022. The hardwood dataset comprised sixteen species, including American elm, cherrybark oak, Nuttall oak, shagbark hickory, Shumard oak, swamp chestnut oak, water oak, willow oak, ash, black gum, deerberry, leatherwood, red maple, post oak, willow, and winged elm. Scientific names and authority of the listed species are shown in Table \ref{tab1}. Nuttall oak, water oak, and Shumard oak were seedlings aged one to three years old, while the others were 30-50 years old when their leaves were collected and used. To prepare the leaves for stomatal peels, we followed the methods described by Hilu and Randall (1984). First, we carefully dried any moisture on the surface of the leaves using paper towels. We applied clear nail polish to 4-6 locations on the abaxial epidermis of the leaves. After about 5-8 minutes, we removed the dried nail polish from the leaves, placed it on pre-cleaned microscope slides, and covered it with one or two coverslips.  In total, over 10,000 stomatal images were captured using a compound light microscope (Olympus, Tokyo, Japan) equipped with a digital microscope camera (MU300, AmScope, USA) with a 5 mm lens and a fixed microscope adapter (FMA050, AmScope). The $Populus$ dataset consists of over 3,000 images. It includes 55 genotypes from seven taxa of eastern cottonwood ($Populus$ $deltoides$ (W. Bartram ex Marshall) × $P$. $deltoides$ (D×D), and $Populus$ hybrids such as $P$. $deltoides$ × $P$. $maximowiczii$ (A. Henry) (D×M), $P$. $deltoides$ × $P$. $nigra$ (L.) (D×N), $P$. $deltoides$ × $P$. $trichocarpa$ (Torr. \& A. Gray ex Hook.) (D×T), $P$. $trichocarpa$ × $P$. $deltoides$ (T×D), and $P$. $trichocarpa$ × $P$. $maximowiczii$ (T×M).

\begin{table*}[t]
\centering

\caption{Plant species used for this study (checked based on Integrated Taxonomic Information System (ITIS, www.itis.gov)).\label{tab1}}
\centering
\tabcolsep=0pt%%
% \usepackage{multirow}
\begin{tabular}{llcrl}
\midrule
Datasets                    & Taxa  & Common names       & Scientific names and authorities           \\
\midrule
\multirow{16}{*}{Hardwoods} & NA    & Ash                  & $Fraxinus$ $caroliniana$ Miller                \\
                            & NA    & Black gum            & $Nyssa$ $sylvatica$ Marshall                   \\
                            & NA    & Deerberry            & $Vaccinium$ $stamineum$ Linneaus               \\
                            & NA    & Leatherwood          & $Dirca$ $palustris$ L.                         \\
                            & NA    & American elm         & $Ulmus$ $americana$ Planch                     \\
                            & NA    & Cherrybark oak       & $Quercus$ $pagoda$ Raf.                        \\
                            & NA    & Red maple            & $Acer$ $rubrum$ L.                             \\
                            & NA    & Nuttall oak          & $Quercus$ $texana$ Buckley                     \\
                            & NA    & Post oak             & $Quercus$ $stellata$ Wangenh.                  \\
                            & NA    & Shagbark hickory     & $Carya$ $ovata$ (Mill.) K. Koch                \\
                            & NA    & Shumard oak          & $Quercus$ $shumardii$ Buckley                  \\
                            & NA    & Swamp chestnut Oak   & $Quercus$ $michauxii$ Nutt.                    \\
                            & NA    & Water oak            & $Quercus$ $nigra$ L.                           \\
                            & NA    & Willow               & $Salix$ L.                                   \\
                            & NA    & Willow oak           & $Quercus$ $phellos$ L.                         \\
                            & NA    & Winged elm           & $Ulmus$ $alata$ Michx.                         \\
$Populus$                     & D×D   & Eastern cottonwood   & $Populus$ $deltoides$ × $Populus$ $deltoides$      \\
                            & D×M   & Poplar hybrids       & $Populus$ $deltoides$ × $Populus$ $maximowiczii$  \\
                            & D×N   & Poplar hybrids       & $Populus$ $deltoides$ × $Populus$ $nigra$          \\
                            & D×N×M & Poplar hybrids       & $Populus$ $deltoides$ × $nigra$ × $maximowiczii$   \\
                            & D×T   & Poplar hybrids       & $Populus$ $deltoides$ × $Populus$ $trichocarpa$    \\
                            & T×D   & Poplar hybrids       & $Populus$ $trichocarpa$ × $Populus$ $deltoides$    \\
                            & T×M   & Poplar hybrids       & $Populus$ $trichocarpa$ × $Populus$ $maximowiczii$\\
\midrule
\end{tabular}
%\begin{tablenotes}%
%\item Note: This is an example of table footnote this is an example of table footnote this is an example of table footnote this is an example of~table footnote this is an example of table footnote
%\item[$^{1}$] Example for a first table footnote.
%\item[$^{2}$] Example for a second table footnote.\vspace*{6pt}
%\end{tablenotes}
\end{table*}

We collected one fully expanded fresh leaf of $Populus$ after measuring selected trees' photosynthetic $\mathrm{CO}_2$ response curves ($A/Ci$) from June to August 2020-2022. The leaves were then placed in labeled plastic bags and stored in a cooler to be transported to the laboratory, where they were kept in a 4°C refrigerator until further preprocessing. As for hardwood leaves, to obtain higher-quality images, we used a 10X upper eyepiece and either an X20 or X40 magnification lens to capture three to ten images per leaf, selecting the appropriate magnification based on the size of the leaf stomata. 

\subsection{Leaf functional traits and tree growth measurement}\label{subsec2}

During the growing season in either June or July, we used a Li-Cor 6400 XT photosynthesis system to measure the assimilation rates, transpiration rates, and stomatal conductance ($g_{s}$) of all taxa of $Populus$, as well as $\mathrm{CO}_2$ response curves (A/Ci curves) to determine their intrinsic photosynthetic capacities. Each curve consists of nine steps of external $\mathrm{CO}_2$ concentrations set in succession in the following order: 400, 300, 200, 100, 50, 400, 400, 600, and 800 ppm. After measuring, leaves were collected, scanned, and measured for specific leaf area (SLA) by drying and weighing them. Collected raw data was pre-processed and calculated using the methods and equations described in Sharkey et al. (2007). Maximum carboxylation rate allowed by ribulose 1·5-bisphosphate carboxylase/oxygenase (RuBisCO) ($V_{cmax}$), rate of photosynthetic electron transport (based on NADPH requirement) ($J_{max}$), triose phosphate use ($TPU$), day respiration ($R_{day}$), and mesophyll conductance ($g_{m}$) were calculated.  Leaf intrinsic water use efficiency (iWUE) and instantaneous water use efficiency (WUE) were calculated using photosynthesis/leaf stomatal conductance and photosynthesis/transpiration, respectively.

We measured leaf area index (LAI) monthly using the LAI-2000C console paired with two wands for each $Populus$ tree, one for above and one for below measurements. The above measurement wand was set on a tripod to auto-log in 10-second intervals, while the below measurement was taken manually facing the same direction as the above wand at around 60 cm above the ground where in relation to the tree crown. We used a 45° view cap to obtain single tree LAI and conducted scattering measurements based on weather conditions to reduce scattering effects. Data were pre-analyzed using FV2200 software for further analyses. Monthly measured LAI were summed to be used as predict variables for productivity estimation. Finally, at the end of the growing season, we measured tree height and diameter at breast height (DBH) using poles or a laser rangefinder and diameter tape and used the collected measurements in an allometric equation to predict dry biomass of $Populus$. 

\subsection{StoManager1 model training and validation}\label{subsec3}

\begin{figure*}[!t]%
\centering
\includegraphics[width=\textwidth, angle=0]{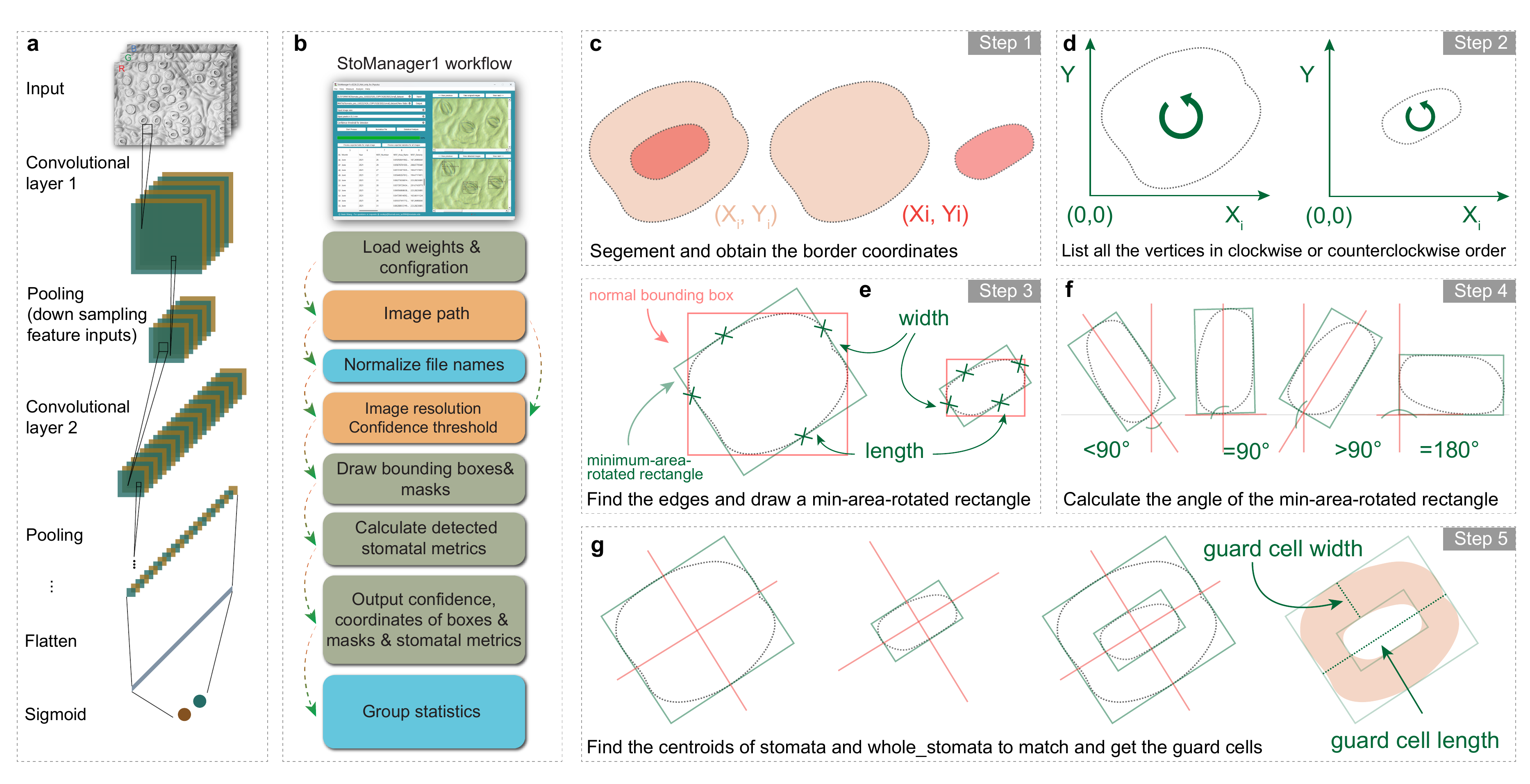}
\caption{Schematic diagrams of a) model training processes, b) detection workflow, and segmentation model pipeline for stomatal metrics measurement (c, d, e, f, g).}\label{fig1}
\end{figure*}

Bounding box- and segmentation-based models were used for StoManager1. Specifically, bounding box-based models take bounding box labels of interested objects for training and generate bounding box outputs for detected objects. Segmentation-based models take mask labels of interested objects for training and generate both bounding boxes and segmentation masks for detected objects. We conducted stratified random sampling to select 265 microscope images from the $Populus dataset$ and 735 images from the hardwood dataset for YOLO bounding box-based detection model training. For the YOLOv8 segmentation model, we used 104 microscope images with 76 images from the hardwood dataset and 28 images from $Populus$ dataset. Two classes, stomata aperture and stomata including guard cells were defined as “stomata” and “whole\textunderscore stomata” for labeling and training, respectively, and labeled using LabelImg and Roboflow for StoManager1 model training (Tzutalin, 2015). StoManager1 models were trained on a Windows Desktop in the Python environment. We cloned and compiled Darknet, and downloaded and applied pre-trained weights from the Darknet website for the convolutional layers (Bochkovskiy et al., 2020; Wang et al., 2021). In the context of convolutional neural networks (CNNs), Darknet generally refers to a specific type of CNN architecture that is designed to be very deep and efficient. The Darknet architecture was first introduced in the YOLO object detection system, which uses a single CNN to detect objects in an image in real-time. The Darknet architecture is characterized by its use of "shortcut connections" that allow information to flow more directly between different layers of the network. This allows the network to be deeper while still maintaining good performance. Detailed training procedures and testing configuration parameters can be found in Fig. \ref{fig1} and a GitHub repository https://github.com/JiaxinWang123/StoManager.git. Pretrained weights darknet53.conv.74 were used for our StoManager1 model training.

Leaf stomatal area and orientation were manually measured using ImageJ (Schindelin et al., 2015) to enable StoManager1 to simultaneously detect and analyze stomatal area and orientation. In this study, the stomatal area refers to either the area of “stomata” of “whole\textunderscore stomata” defined as the area of the individual stoma aperture and stomatal aperture with guard cells respectively, while orientation was defined as the angle of the individual stoma based on a horizontal line of the stomatal image that was captured. It was impossible to ensure that all images were taken at the same horizontal level, so the aim was not to specify the orientation of each stoma individually. Instead, the goal was to quantify the variance in leaf stomatal orientation. A minor orientation variance would indicate that the stomata were arranged more orderly, while a larger orientation variance would suggest that the stomata were arranged more disorderly, and those variations would significantly affect leaf photosynthetic assimilation, maximum leaf stomatal conductance, and water use efficiency by regulating the space, pathway, and patterning between the stomata and leaf veins (Croxdale, 2000; Dow et al., 2014). Guard cell width was defined as the perpendicular distance between the lengths of the minimum-area-rotated rectangles of the guard cells and “stomata”. Guard cell length was defined as the length of the minimum-area-rotated rectangle of the guard cells (Fig. \ref{fig1}).

We used various methods to measure or predict relevant metrics for “stomata” and “whole\textunderscore stomata.” We masked the segmentation area for segmentation-based models and directly measured their properties for further calculation. For bounding box-based models, we extracted the height and width of the boxes and used them, along with their product, ratio, or squares, to fit linear regression models to predict the metrics of interest. To obtain ground-truth values, we measured metrics using ImageJ (Schindelin et al., 2015).

For model validation, we employed precision and recall metrics. Precision and recall are two commonly used metrics for evaluating the performance of image detection systems (Padilla et al., 2020). Precision is the ratio of true positive detections to the total number of detections made by the model. In other words, it measures the model's accuracy by looking at how many of the detections made were correct. A high precision score indicates that the system has a low false positive rate, i.e., it is good at correctly identifying the objects in the image. Recall is the ratio of true positive detections to the total number of objects in the image. It measures the completeness of the system by looking at how many objects in the image were correctly detected. A high recall score indicates that the system has a low false negative rate, i.e., it is good at not missing any objects in the image. Both precision and recall are essential metrics for image detection, as they help to evaluate the system's performance from different perspectives. We combined the single object detection confidence level (i.e., the likelihood that an object belongs to a certain class) and the overlap of ground-truth bounding box $B_{gt}$ and the detected bounding box $B_{p}$ and applied intersection over union (IOU) to evaluate the ratio of the intersection area to the union area of $B_{gt}$ and $B_{p}$ (Jaccard, 1901).

We used mean average precision with an IOU threshold of 0.50 (mAP@0.50) and 0.50-0.90 to assess our model's performance. Due to our scenario's lack of true negatives, we opted not to utilize the confusion matrix and other metrics calculated based on true negative detection, as they are not applicable (Padilla et al., 2021). Details about calculating precision and recall can be found in Padilla et al. (2021). To validate our bounding box-based model, we chose a set of 150 images that were not previously seen from the $Populus$ dataset, another set of 100 unseen images from the hardwood dataset, and an additional 100 images of American holly ($Ilex$ $opaca$, Aiton), which had not been encountered before. We manually labeled them as either "whole\textunderscore stomata" or "stomata" to evaluate the accuracy of StoManager1's detection. Additionally, we randomly selected 588 images from nine species in the hardwood dataset and manually labeled and measured their stomatal area and orientation based on a horizontal line using ImageJ to validate StoManager1's ability to measure stomatal properties for unseen hardwood images accurately. For the segmentation model, validation was performed during the training process using 30 images with 18 images from the hardwood dataset and 12 images from the $Populus$ dataset.

\subsection{Stomatal parameter definition and measurement}\label{subsec4}

The commonly measured stomatal metrics, such as stomatal numbers and density, can be easily obtained by extracting the number of “whole\textunderscore stomata” from the YOLOv8 outputs. To be specific, stomatal density can be calculated using Eq. 1:

\begin{equation}
D=\ \frac{{100\times N\times P}^2}{A_{img}}\label{eq1}
\end{equation}

where $D$ is the stomatal density (stomata $mm^{-2}$), $N$ is the total number of “whole\textunderscore stomata” within the tested image, $P$ is the number of pixels per 0.1 mm line, 100 is used to convert 0.01 $mm^2$ to 1.0 $mm^2$, and the $A_{img}$ is the area of the input image ($pixel^2$).

\subsubsection{Empirical algorithms}\label{subsubsec2}

To estimate specific stomatal and guard cell metrics, linear regression models were trained using various features extracted from bounding box-based model outputs, such as bounding box height and width. These metrics include “stomata” and “whole\textunderscore stomata” area, “stomata” and “whole\textunderscore stomata” orientation, variances in “stomata” and “whole\textunderscore stomata” area and orientation, guard cell width, and guard cell length. To estimate the “stomata” area and “whole\textunderscore stomata” area, we used the product of width and height of the YOLO extracted normal bounding boxes as the independent variables to fit simple linear regression models, as shown in Eq. 2 and 3:

\begin{equation}
A_{st}=\ \left(\frac{w\times h+116.08}{1.7684}\right)\times\frac{{10}^4}{P^2}\label{eq1}
\end{equation}

\begin{equation}
A_{wst}=\ \left(w\times h\times0.6878+806\right)\times\ \frac{{10}^4}{P^2} \label{eq1}
\end{equation}

where $A_{st}$ represents the area of “stomata” ( \textmu $m^2$), $w$ and $h$ represent the width and height of corresponding boxes of “stomata” and “whole\textunderscore stomata”, and $A_{wst}$ represents the area of “whole\textunderscore stomata” (\textmu $m^2$). $P$ is the number of pixels per 0.1 mm line.

We used the ratio of bounding box width and height to estimate the orientation of “stomata” and “whole\textunderscore stomata” (Eq. 4):

\begin{equation}
R=\ln{\left(\frac{w}{h}\right)}+44.5222 \label{eq1}
\end{equation}

where R is the orientation of “stomata” and “whole\textunderscore stomata” which is defined as the angle of the individual “stoma”/ “whole\textunderscore stoma” based on the horizontal line of the stomatal image.

We use bounding box width and height to estimate the guard cell width and length using the following equations (Eq. 5, 6):

\begin{equation}
W_{gc}=\frac{\left(0.12\times w_1-0.10\times h_1+0.09\times w_2+0.34\times h_2+9.27\right)\times P}{100}\label{eq1}
\end{equation}

\begin{equation}
L_{gc}=\frac{\left(0.28\times w_1+0.29\times h_1+0.16\times w_2+0.13\times h_2+25.07\right)\times P}{100}\label{eq1}
\end{equation}

where $W_{gc}$ (\textmu m) is the guard cell width, $L_{gc}$ (\textmu m) is the guard cell length, which are shown in Fig. 4, $w_1$ and $h_1$ are the width and height of “whole\textunderscore stomata” bounding box, and $w_2$ and $h_2$ are the width and height of the “stomata” bounding box. $P$ is the number of pixels per 0.1 mm line.

\subsubsection{Theoretical algorithms}\label{subsubsec2}

We built a framework using geometrical and mathematical algorithms to measure over 30 stomatal and guard cell metrics such as “stomata” area, “whole\textunderscore stomata” area, “stomata” length, “stomata” width, “stomata” orientation (angle), guard cell length, guard cell width, guard cell area, and “stomata” area/guard cell area ratio. More metrics are listed in Table \ref{tab2}. The framework is shown in Fig. \ref{fig1} and listed below:

\begin{table*}[t]
\centering

\caption{Metrics of “stomata”, “whole\textunderscore stomata”, and guard cell measured by StoManager1.\label{tab2}}
\centering
\tabcolsep=0pt%%
% \usepackage{multirow}

\begin{tabular}{lcrl}
\midrule
Ouput variables        & Defination                            & Unit       &  \\
\midrule
ori\_img\_shape         & original image shape, e.g., (1024, 768)                     & NA          &  \\
class\_wst             & class of whole\_stomata, e.g., "1"    & NA         &  \\
number\_wst            & total number of whole\_stomata        & NA         &  \\
index\_wst             & index of whole\_stomata               & NA         &  \\
box\_w\_wst            & bounding box width of whole\_stomata  & pixels     &  \\
box\_h\_wst            & bounding box height of whole\_stomata & pixels     &  \\
area\_wst              & area of whole\_stomata                & $\mu m^2$        &  \\
width\_wst             & width of whole\_stomata               & $\mu m$         &  \\
length\_wst            & length of whole\_stomata              & $\mu m$         &  \\
var\_area\_wst         & variance of whole\_stomata area       & NA         &  \\
var\_width\_wst        & variance of whole\_stomata width      & $\mu m^2$        &  \\
var\_length\_wst       & variance of whole\_stomata length     & $\mu m^2$        &  \\
centroid\_wst          & centroid of whole\_stomata            & NA         &  \\
class\_st              & class of whole\_stomata, e.g., "0"    & NA         &  \\
number\_st             & total number of stomata               & NA         &  \\
index\_st              & index of stomata                      & NA         &  \\
box\_w\_st             & bounding box width of stomata         & pixels     &  \\
box\_h\_st             & bounding box height of stomata        & pixels     &  \\
area\_st               & area of stomata                       & $\mu m^2$        &  \\
width\_st              & width of stomata                      & $\mu m$         &  \\
length\_st             & length of stomata                     & $\mu m$         &  \\
var\_area\_st          & variance of stomata area              & NA         &  \\
var\_width\_st         & variance of stomata width             & $\mu m^2$        &  \\
var\_length\_st        & variance of stomata length            & $\mu m^2$        &  \\
centroid\_st           & centroid of stomata                   & NA         &  \\
guardCell\_length      & guard cell length                     & $\mu m$         &  \\
guardCell\_width       & guard cell width                      & $\mu m$         &  \\
guardCell\_area        & guard cell area                       & NA         &  \\
guardCell\_angle       & orientation of guard cell             & degree &  \\
var\_angle             & variance of stomatal orientation      & degree &  \\
var\_width\_guardCell  & variance of guard cell width          & NA         &  \\
var\_length\_guardCell & variance of guard cell length         & NA         &  \\
var\_area\_guardCell   & variance of guard cell area           & NA         &  \\
wst\_density            & whole\_stomata density in each image                        & stomata/$\mu m^2$ &  \\
ratio\_area\_st\_to\_gc & ratio of stomata area to guard cell area                    & $\mu m^2$/$\mu m^2$     &  \\
ratio\_area\_to\_img    & ratio of the sum of all whole\_stomata area to   image area & $\mu m^2$/$\mu m^2$    &  \\
SEve                   & stomatal evenness index               & NA         &  \\
SDiv                   & stomatal divergence index             & NA         &  \\
SAgg                   & stomatal aggregation index            & NA         & \\
\midrule
\end{tabular}
%\begin{tablenotes}%
%\item Note: This is an example of table footnote this is an example of table footnote this is an example of table footnote this is an example of~table footnote this is an example of table footnote
%\item[$^{1}$] Example for a first table footnote.
%\item[$^{2}$] Example for a second table footnote.\vspace*{6pt}
%\end{tablenotes}
\end{table*}

1. We used the YOLO segmentation algorithm (YOLOv8x-seg) to obtain the segmentation of “whole\textunderscore stomata” and “stomata” and extracted the coordinates of segments. The segmentation model gives the coordinates of vertices of the masks, which can be extracted from the prediction results.

2. The Surveyor's Formula was utilized to compute the segmented areas (Braden, 1986):

\begin{equation}
A_s=\frac{\left(\frac{1}{2}\left\{\left|\begin{matrix}x_0&x_1\\y_0&y_1\\\end{matrix}\right|+\left|\begin{matrix}x_1&x_2\\y_1&y_2\\\end{matrix}\right|+\ldots+\left|\begin{matrix}x_{n-2}&x_{n-1}\\y_{n-2}&y_{n-1}\\\end{matrix}\right|+\left|\begin{matrix}x_{n-1}&x_0\\y_{n-1}&y_0\\\end{matrix}\right|\right\}\right)\times{10}^4}{P^2}\label{eq1}
\end{equation}

where $A_s$ represents the area (\textmu $m^2$) of “stomata” or “whole\textunderscore stomata”, $x_0$, $y_0$, $x_1$, $y_1$, $x_{n-2}$, $y_{n-2}$, $x_{n-1}$, and $y_{n-1}$ are the coordinates of vertices of the segment, and $P$ represents the number of pixels per 0.1 mm line.

3. We found the edges and drew a minimum-area-rotated rectangle to obtain the width and length based on the methods described by Freeman and Shapira (1975).

4. We used the length of the minimum-area-rotated rectangle to calculate the orientation (angle based on the horizontal line, Fig. \ref{fig1}).

5. We found the centroids of “whole\textunderscore stomata” and “stomata” and matched them to extract guard cells. Specifically, we obtained guard cell masks by subtracting “stomata” masks using “whole\textunderscore stomata” masks. \bigskip
\bigskip

\bigskip

\subsubsection{Stomatal indices}\label{subsubsec2}

Stomata require sufficient space to work properly, which is ensured by the "one cell spacing rule" (Parlange \& Waggoner, 1970; Franks \& Casson, 2014). Here, we calculated stomatal indices, such as stomatal evenness (${SE}_{ve}$), divergence ($SD_{iv}$), and aggregation (${SA}_{gg}$) index as described by Liu et al. (2021).

Following equations from Eq. 8 to Eq. 18, ${SE}_{ve}$, $SD_{iv}$, and ${SA}_{gg}$ were calculated. Stomatal evenness refers to the consistency of stomatal distribution across a leaf's surface. During this procedure, the minimum spanning tree (MST) connects all stomata by minimizing the total branch lengths (distance between stomata) and yields $N-1$ branches for $N$ stomata. Initially, each branch's length is divided by the total length of all branches to determine the partial distance ($PD$) for every branch.

\begin{equation}
{PD}_l=\ \frac{D_l}{\sum_{l=1}^{N-1}D_l}\label{eq1}
\end{equation}

where $D$ represents the Euclidean distance between stomata, with branch $l$ containing the involved stomata. 

\begin{equation}
{SE}_{ve}=\frac{\sum_{l=1}^{N-1}{min\left({PD}_l,\frac{1}{N-1}\right)-\frac{1}{N-1}}}{1-\frac{1}{N-1}}\label{eq1}
\end{equation}

By subtracting $1/( N-1)$ from both the numerator and denominator, the analysis accounts for the fact that there will always be at least one $PD_l$ value less than or equal to $1/( N-1)$, irrespective of the N value. Consequently, ${SE}_{ve}$ is a unitless metric with a range between 0 and 1. When all $PD_l$ values are equal to $1/( N-1)$, ${SE}_{ve}$ reaches a value of 1. To establish an MST and subsequently estimate ${SE}_{ve}$, each image must contain more than three stomata.

Stomatal divergence refers to the variation in stomatal distribution across a leaf's surface. To begin, we calculated the coordinates of the center of gravity "G" for the $N$ stomata present in the image.

\begin{equation}
G=\ \frac{1}{N}x_i\label{eq1}
\end{equation}

where $x_i$ is the coordinates of the $i$th stoma.

For each of the $N$ stomata, the Euclidean distance to the center of gravity was determined as Eq. 11.

\begin{equation}
dG_i=\sqrt{\sum{{(x}_i-G)}^2}\label{eq1}
\end{equation}

We calculated the mean distance of the $N$ stomata to the center of gravity using Eq. 12.

\begin{equation}
\bar{dG}=\frac{1}{N}\sum_{i=1}^{N}{dG}_i\label{eq1}
\end{equation}

The sum of deviances ($\Delta d$) and absolute ($\Delta  | d |$) distances from the center of gravity, respectively, across the stomata were calculated as Eq. 13 and 14 to calculate $SD_{iv}$ as Eq. 15.

\begin{equation}
 \Delta d =  \sum_{i=1}^N (dG_i- \overline{dG} )\label{eq1}
\end{equation}

\begin{equation}
 \Delta  | d | = \sum_{i=1}^N  | dG_i- \overline{dG} | \label{eq1}
\end{equation}

\begin{equation}
SD_{iv} =  \frac{ \Delta d+ \overline{dG} }{ \Delta  | d | + \overline{dG} } \label{eq1}
\end{equation}

The stomatal aggregation index measures the extent to which the observed stomatal distribution deviates from a random expectation concerning the distance to the nearest neighbor. To calculate the nearest-neighbor distance for each stoma, the formula developed by Clark and Evans (1954) was followed (Eq. 16). ${SA}_{gg}$ were calculated as the ratio of the observed nearest-neighbor distance to the theoretical nearest-neighbor distance as Eq. 17 and 18.

\begin{equation}
d=\ \frac{1}{{2(\frac{N}{Area})}^{0.5}}\label{eq1}
\end{equation}

\begin{equation}
\bar{d}=\frac{1}{N}\sum_{i=1}^{N}d_i\label{eq1}
\end{equation}

\begin{equation}
{SA}_{gg}=\frac{\bar{d}}{d}\label{eq1}
\end{equation}

${SA}_{gg}$ values range between 0 and 2.15. These values signify different patterns of distribution: perfectly uniform (${SA}_{gg}$ value $>$ 1), random (${SA}_{gg}$ value = 1), and entirely aggregated (${SA}_{gg}$ value $<$ 1).

\subsection{Hardware, software, and statistics}\label{subsec2}

The experimental environment of this paper is as follows: the Windows 11 operating system (CUDA 11.7 version, for GPU acceleration) and Anaconda 3. The GPU is MSI GeForce RTX 3080Ti 12GB, and the CPU is Intel Core i7 12700KF. Open-source languages Python (3.10.8 64-bit) and R (version 4.2.2) were used for coding, application building, data analysis, and visualization (Van Rossum \& Drake, 2009; R Core Team, 2022). For group analysis to calculate the mean, median, minimum, and maximum value of stomatal metrics in each image, we removed outliers based on $97.5^{th}$ percentile point before calculating. These outliers were introduced because of the objects detected at the edges of images. The median, mean, and variance were computed using the NumPy module in Python. Each image has a relevant output of size of each stoma or total stomatal area in the image, density, and orientation measurement. We did principal component analysis (PCA) to explore the underlying relationships between measured stomatal metrics (108 metrics) and plant physiological and growth performance. Principal component analysis (PCA) results suggested that most of the StoManager1 measured stomatal metrics exhibited substantial correlation with $Populus$' physiological and growth performance. We used the built-in R functions prcomp() for the PCA and the package “factoextra” to plot the output. Correlations were checked using simple linear regression and the correlation test in R. Before conducting ANOVA and linear regression, we checked for independence, normality, and homogeneity of variance assumptions. If any of these assumptions were not met, we applied a log-transformation to the variable. We fitted multiple and weighted multiple linear regression models to explore the potential of using StoManager1-measured metrics and leaf functional traits to predict $Populus$ productivity. RMSE, RMSE \%, weighted RMSE\%, $R^2$, and Akaike information criterion (AIC) were used for the model's evaluation and selection (Hu, 2007). 

\section{Results}\label{sec3}

\subsection{StoManager1 detection performance}\label{subsec2}

\begin{figure*}[!t]%
\centering
\includegraphics[width=\textwidth, angle=0]{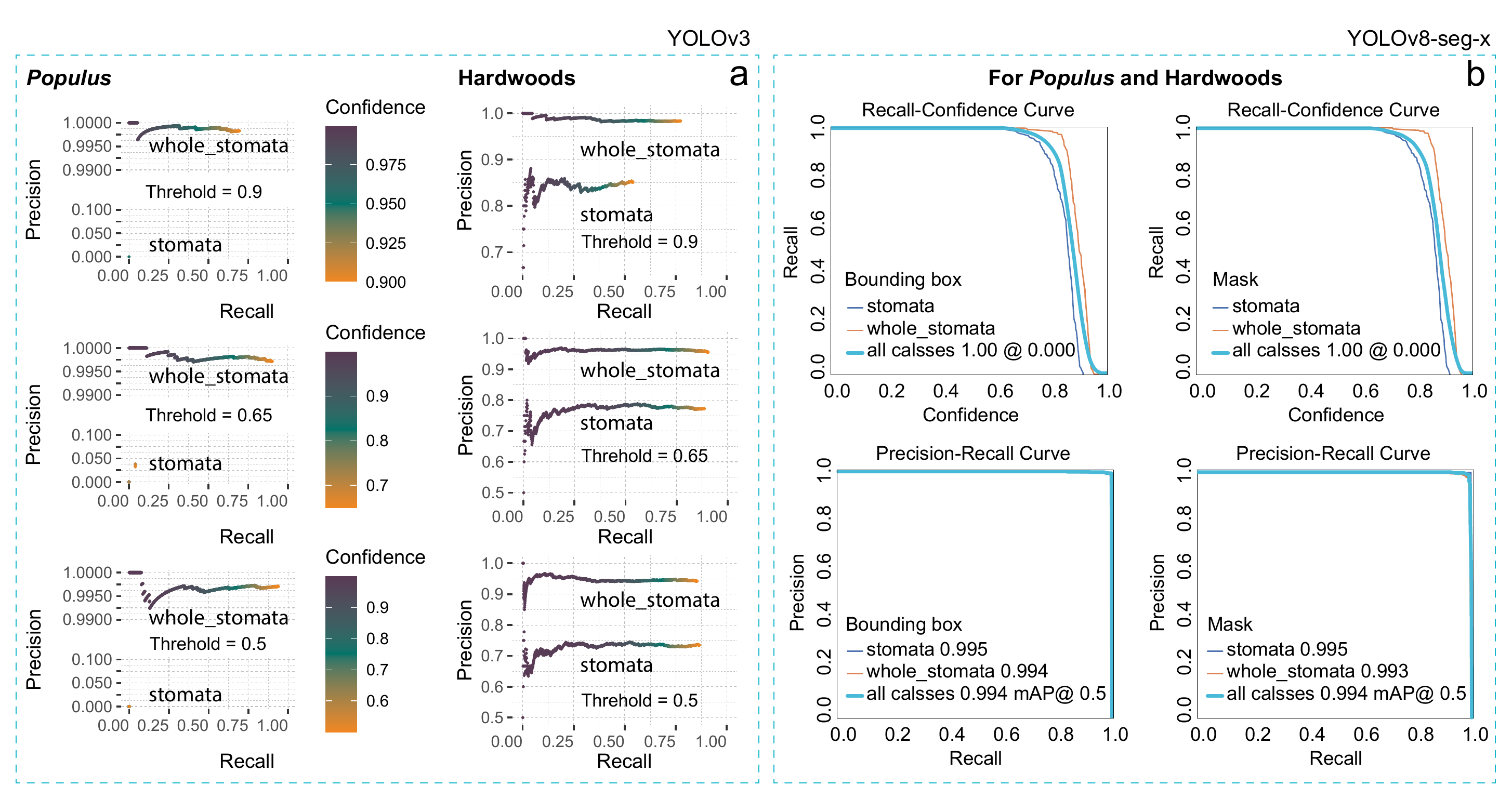}
\caption{Precision × recall curves and recall × confidence curves of a) bounding box- and b) segmentation-based models in StoManager1 for hardwood and $Populus$ datasets.}\label{fig2}
\end{figure*}

\begin{figure*}[!t]%
\centering
\includegraphics[width=\textwidth, angle=0]{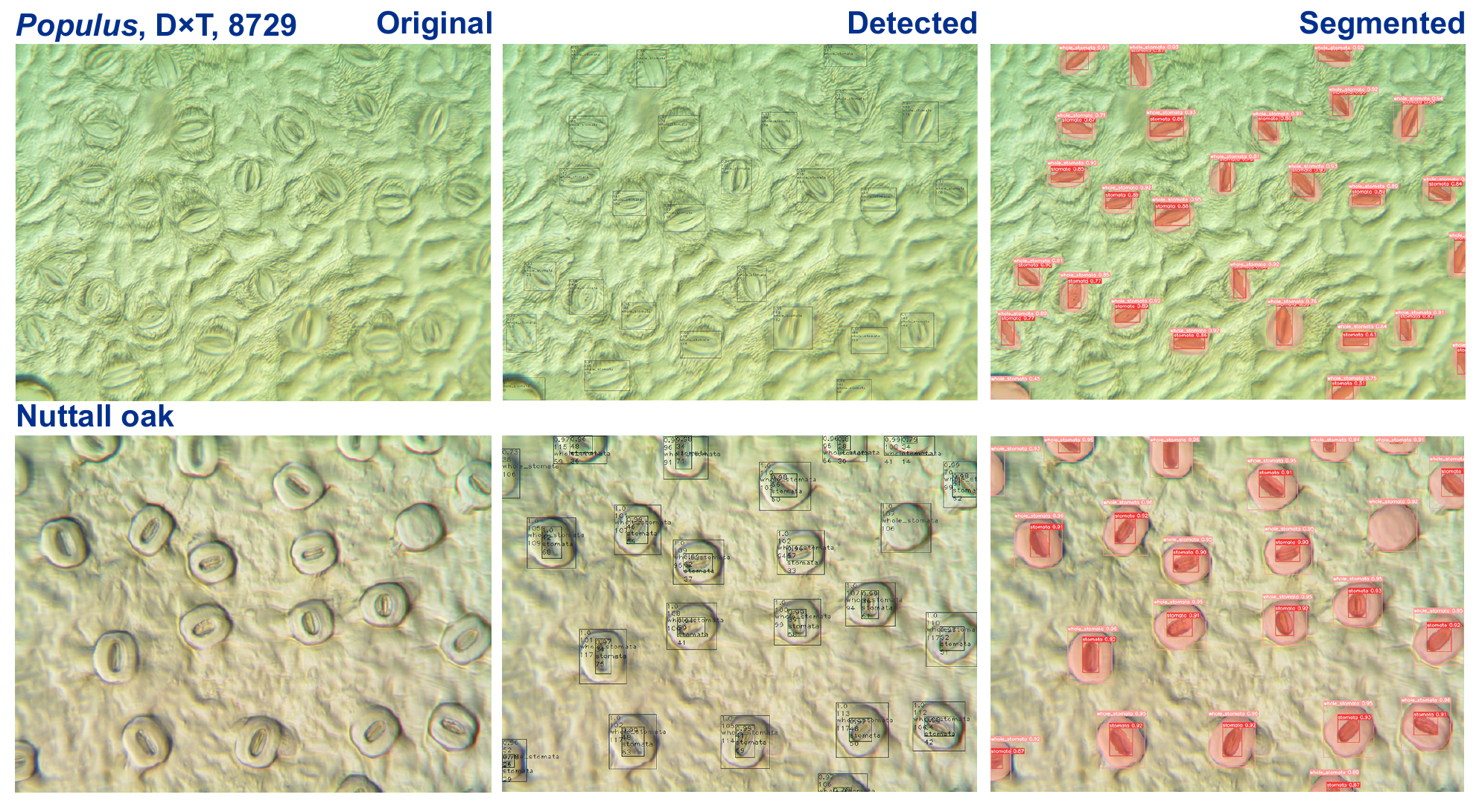}
\caption{Examples of model detection and segmentation.}\label{fig3}
\end{figure*}

StoManager1 can detect “stomata” and “whole\textunderscore stomata” in all images of various taxa and genotypes from the Populus dataset and species from the hardwood dataset (Fig. \ref{fig2}). The mAP was calculated based on the precision of the two classes (“stomata” and “whole\textunderscore stomata”), and recall was used to evaluate the model's performance. The results indicated that StoManager1 demonstrated high precision in detecting “whole\textunderscore stomata” in both the $Populus$ and hardwood datasets. The mAPs were influenced by the confidence threshold and the precision for each class (Fig. \ref{fig2}. Generally, StoManager1 exhibited higher mAPs for the hardwood dataset (61-78\% when confidence thresholds for detection set as 0.5, 0.65, and 0.9) than the $Populus$ dataset (36-47\%, with precision for class 'stomata' close to zero). When the confidence threshold was set higher, precision increased, but recall decreased, and when it was set lower, precision decreased, but recall increased. StoManager1 was found to be more effective in detecting “stomata” in the hardwood dataset than in the $Populus$ dataset. Specifically, the precision of “stomata” detection for the $Populus$ dataset ranged from 0 to 0.5, whereas for the hardwood dataset, it ranged from 0.5 to 1.0 when the confidence thresholds were set from 0.5 to 0.9. There was notable diversity in the quality of the micrographs we observed, with specific images being unevenly focused, leading to blurring and structural incompleteness in some objects (Fig. \ref{fig3}). 

\subsection{StoManager1 measuring performance}\label{subsec2}

\begin{figure*}[!t]%
\centering
\includegraphics[width=\textwidth, angle=0]{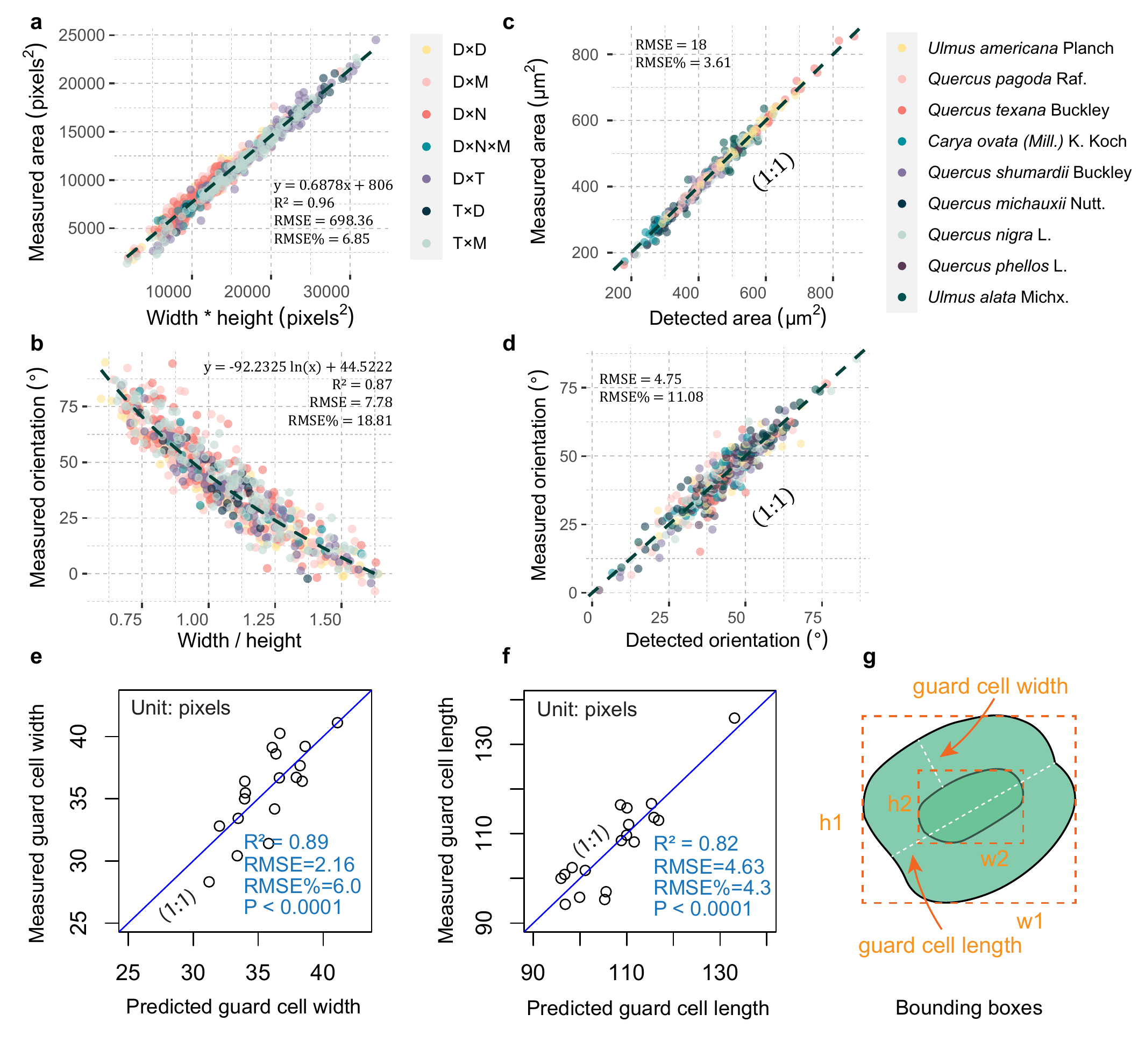}
\caption{Models trained using the $Populus$ and validated using the hardwood dataset for stomatal area and orientation estimation. a) linear regression model fitted using the $Populus$ dataset with the product of the stomatal bounding box's width and height (in $pixels^2$) as the independent variable and measured individual stomatal area (in $pixels^2$).  b) regression of measured individual stomatal orientation and the ratio of stomatal width and height, and c, d) validation of the regression models using 16 hardwood species datasets. e, f, g) bounding box-based model performance in predicting guard cell width and length. D×D = $P.$ $deltoides$ × $P.$ $deltoides$, D×M = $P.$ $deltoides$ × $P. maximowiczii$, D×N = $P.$ $deltoides$ × $P.$ $nigra$, D×T = $P.$ $deltoides$ × $P.$ $trichocarpa$, T×M = $P.$ $trichocarpa$ × $P.$ $maximowiczii$, and D×N×M = D×N × $P.$ $maximowiczii$.}\label{fig4}
\end{figure*}

Our findings indicated that the bounding box-based, model-measured leaf stomatal area is linearly related to the product of width and height ($R^2$ = 0.9606, RMSE\% = 6.85, P $<$ 0.0001), and the measured leaf stomatal orientation is logarithmically linked to the ratio of detected width and height ($R^2$ = 0.868, RMSE\% = 18.81, P $<$ 0.0001) (Fig. \ref{fig4}a, b). Moreover, we applied these models to determine the stomatal properties of images from the hardwood dataset. The outcomes revealed that the models have low losses in measuring stomatal area and orientation, with RMSE values of 18 $\mu m^2$ (RMSE\% = 3.61) and 4.75° (RMSE\% = 11.08) for stomatal area and orientation, respectively (Fig. \ref{fig4}c, d). Weighted multiple linear regression models showed substantial ability in predicting guard cell width ($R^2$ = 0.89, RMSE\% = 6.0, P $<$ 0.0001) and length ($R^2$ = 0.82, RMSE\% = 4.3, P $<$ 0.0001) using bounding box-based model output (Fig. \ref{fig4}e, f, g). The StoManager1 segmentation model showed superior capability in detecting and measuring stomatal metrics than the bounding box-based model. For example, the precision, recall, and mAP@0.50 for all classes of the segmentation model are above 0.994 (Fig. \ref{fig2}b). 

\subsection{Biological applications}\label{subsec2}

\begin{figure*}[!t]%
\centering
\includegraphics[width=\textwidth, angle=0]{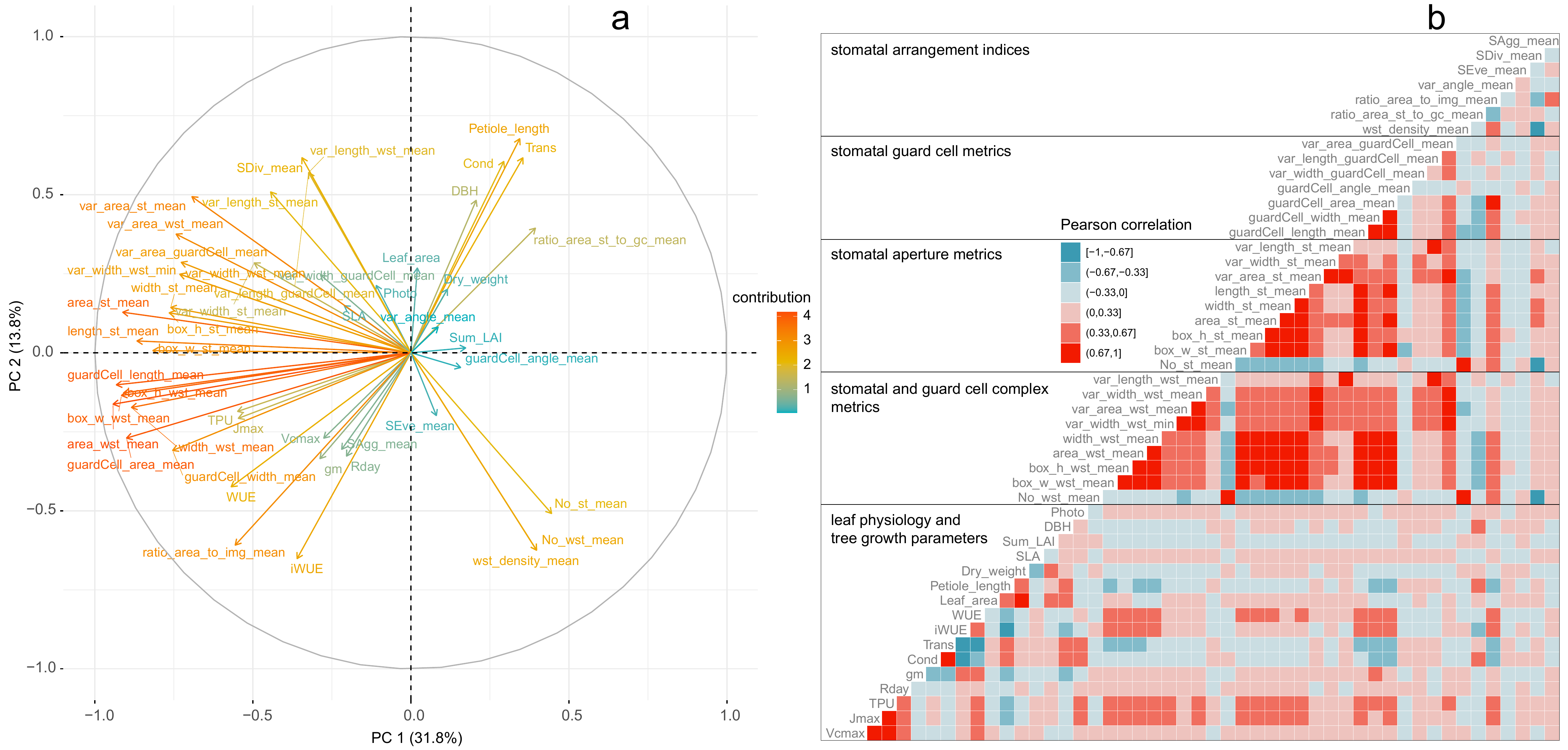}
\caption{Principal component analysis (PCA) and Pearson correlation analysis results of StoManager1 measured metrics, leaf functional traits, and tree growth measurements. $V_{cmax}$ is the maximum carboxylation rate, $J_{max}$ is the maximum rate of photosynthetic electron transport, TPU is triose phosphates, $R_{day}$ is the light respiration rate in absence of photorespiration, $g_m$ is mesophyll conductance, Cond is stomatal conductance, Trans is leaf transpiration rate, iWUE is intrinsic water use efficiency, WUE is instantaneous water use efficiency, SLA is specific leaf area, and Photo is photosynthesis. “wst” is “whole\_stomata” defined as stomatal aperture and guard cell complex, “st” is “stomata” defined as stomatal aperture, “gc” is guard cell, ${SE}_{ve}$ is stomatal evenness index, ${SD}_{iv}$ is stomatal divergence index, and ${SA}_{gg}$ is stomatal aggregation index.}\label{fig5}
\end{figure*}

We investigated the potential biological applications of StoManager1 measured stomatal metrics by examining the underlying correlations between these metrics and the physiological and growth performance of 55 genotypes of $Populus$ (Fig. \ref{fig5}). In general, PC1 and PC2, explained 32\% and 13\% of the total variance of all variables, respectively. Guard cell and “stomata” relevant metrics such as guard cell width, length, area, and “stomata” width, length, and area showed positive correlations with leaf photosynthetic capacity, water use efficiency, and/or negative correlations with leaf $g_s$ and plant growth (Fig. \ref{fig5}). For instance, the mean guard cell length (correlation coefficient = 0.50, p $<$ 0.0001) and area (correlation coefficient = 0.48, p $<$ 0.0001) showed significantly positive correlation with $J_{max}$. The minimum guard cell width (correlation coefficient = 0.24, p $<$ 0.0001), minimum guard cell area (correlation coefficient = 0.38, p $<$ 0.0001), and minimum “whole\textunderscore stomata” area (correlation coefficient = 0.30, p $<$ 0.0001) displayed substantial positive correlation with the leaf iWUE of $Populus$. The mean guard cell width (correlation coefficient = 0.48, p $<$ 0.0001) and mean guard cell area (correlation coefficient = 0.52, p $<$ 0.0001) showed significantly positive correlations with leaf WUE. The median ratio of “stomata” area/guard cell area showed a significantly positive correlation with leaf $g_s$ (correlation coefficient = 0.45, p $<$ 0.0001). The variance in stomatal orientation (angle) showed positive correlations with leaf $g_s$, transpiration, and DBH of $Populus$ (Fig. \ref{fig5}).

Multiple and weighted multiple linear regression models were also fitted using various factors such as petiole length, mean guard cell length, ratio of “stomata” area/guard cell area, sum of monthly leaf area index, and leaf area to explain $Populus$ productivity (Table \ref{tab3}). Multiple linear regression and weighted multiple linear regression models showed good explanatory power, with over 38\% and 78\% of the variance in $Populus$ productivity explained using predicting variables. Multiple linear regression and weighted multiple linear regression models fitted using leaf petiole length, ratio of “stomata” area/guard cell area, minimum guard cell width, sum of monthly leaf area index, and leaf area explained over 55\% and 93\% of the variance in iWUE of $Populus$. The best models for $Populus$ productivity and iWUE prediction are weighted multiple linear regression models, and the RMSE\% and weighted RMSE\% of weighted multiple linear regression model are 39\% and 9\% for productivity and 18\% and 6\% for iWUE, respectively.

\begin{sidewaystable}%[!p]
\caption{Multiple linear regression models fitted using leaf functional traits to predict $Populus$ productivity and leaf intrinsic water use efficiency (iWUE).\label{tab3}}
\tabcolsep=0pt%
\begin{tabular*}{\textwidth}{@{\extracolsep{\fill}}lllccccccccl@{\extracolsep{\fill}}}

\toprule
Model &
  Dependent variable &
  Independent variable &
  Estimate &
  Std. Error &
  t value &
  Pr(\textgreater{}t) &
  Residual standard error &
  Adjusted R-squared &
  F-statistic &
  p-value \\
\midrule
multiple linear regression & Productivity & Intercept                     & -1449.21 & 773.376 & -1.874  & 0.0615            & 1868  & 0.3813 & 90.21 & \textless{}0.0001 \\
                           &              & Petiole\_length               & 618.485  & 40.785  & 15.164  & \textless 0.0001  &       &        &       &                   \\
                           &              & guardCell\_length\_mean       & 26.893   & 10.2    & 2.637   & \textless 0.0001  &       &        &       &                   \\
                           &              & Sum\_LAI                      & 184.749  & 26.13   & 7.07    & \textless{}0.0001 &       &        &       &                   \\
                           &              & Leaf\_area                    & -7.597   & 1.376   & -5.522  & \textless{}0.0001 &       &        &       &                   \\
weighted multiple linear regression &
  Productivity &
  Intercept &
  -1600.59 &
  215.5674 &
  -7.425 &
  \textless{}0.0001 &
  37.97 &
  0.7843 &
  527.4 &
  \textless{}0.0001 \\
                           &              & Petiole\_length               & 627.4302 & 19.2657 & 32.567  & \textless{}0.0001 &       &        &       &                   \\
                           &              & guardCell\_length\_mean       & 25.3246  & 3.0726  & 8.242   & \textless{}0.0001 &       &        &       &                   \\
                           &              & Sum\_LAI                      & 190.1725 & 10.1194 & 18.793  & \textless{}0.0001 &       &        &       &                   \\
                           &              & Leaf\_area                    & -7.0843  & 0.6329  & -11.194 & \textless{}0.0001 &       &        &       &                   \\
multiple linear regression & iWUE         & Intercept                     & 58.585   & 1.437   & 40.763  & \textless{}0.0001 & 7.495 & 0.552  & 143.7 & \textless{}0.0001 \\
                           &              & Petiole\_length               & -3.253   & 0.153   & -21.209 & \textless{}0.0001 &       &        &       &                   \\
                           &              & ratio\_area\_st\_to\_gc\_mean & -1.565   & 0.423   & -3.703  & \textless{}0.0001 &       &        &       &                   \\
                           &              & guardCell\_width\_min         & 1.262    & 0.645   & 1.956   & 0.0509            &       &        &       &                   \\
                           &              & Sum\_LAI                      & -0.284   & 0.1     & -2.831  & 0.0048            &       &        &       &                   \\
                           &              & Leaf\_area                    & 0.04     & 0.005   & 8.142   & \textless{}0.0001 &       &        &       &                   \\
weighted multiple linear regression &
  iWUE &
  Intercept &
  57.445 &
  0.487 &
  117.889 &
  \textless{}0.0001 &
  2.314 &
  0.9299 &
  1536 &
  \textless{}0.0001 \\
                           &              & Petiole\_length               & -3.245   & 0.055   & -59.334 & \textless{}0.0001 &       &        &       &                   \\
                           &              & ratio\_area\_st\_to\_gc\_mean & -1.36    & 0.138   & -9.888  & \textless{}0.0001 &       &        &       &                   \\
                           &              & guardCell\_width\_min         & 1.001    & 0.231   & 4.332   & \textless{}0.0001 &       &        &       &                   \\
                           &              & Sum\_LAI                      & -0.273   & 0.043   & -6.369  & \textless{}0.0001 &       &        &       &                   \\
                           &              & Leaf\_area                    & 0.043    & 0.002   & 20.909  & \textless{}0.0001 &       &        &       &                   \\
\midrule
\end{tabular*}

\end{sidewaystable}

\section{Discussion}\label{sec4}

\subsection{StoManager1 performance}\label{subsec2}

StoManager1 models demonstrated high precision and recall scores, indicating their effectiveness in detecting and counting “whole\textunderscore stomata” on both the $Populus$ and hardwood datasets. However, the bounding box-based model exhibited lower precision and recall for detecting stomata (without guard cells) on the $Populus$ dataset mainly because most “stomata” in the $Populus$ images were closed or too small to be detected using bounding box-based model. The StoManager1 bounding box-based model had a higher mean average precision (mAP) for detecting both stomata and “whole\textunderscore stomata” for the hardwood dataset compared with the $Populus$ dataset, which was attributed to the lower precision in detecting stomata (without guard cells) for $Populus$. We observed that most of the bounding box-based model detected objects in the $Populus$ dataset were closed “whole\textunderscore stomata”, which was due to the lower number of “stomata” caused by the closure of leaf stomata during or after sampling. The StoManager1 segmentation-based model showed superiority in detecting and measuring stomatal metrics, which may be attributed to the usage of theoretical algorithms. Despite the weakness of the bounding box-based model in measuring stomatal metrics compared with the segmentation model, it is helpful in processing low quality images, in which the edges of stomata and guard cells are not fully visible.

Like other studies, we also observed a tradeoff between precision and recall, with lower precision resulting in higher recall, which means more objects can be detected. The recall scores decreased for both “whole\textunderscore stomata” and “stomata” classes due to blurred and structurally incomplete objects within the images. To improve recall scores for blurred and structurally incomplete objects in the “whole\textunderscore stomata” and “stomata” classes, we might need to consider improving image quality or using image quality enhancing techniques when taking images; applying preprocessing techniques such as filtering, contrast enhancement, and image normalization; implementing a more sophisticated image segmentation algorithm; augmenting the dataset to include more diverse images; and fine-tuning the StoManager1 model with the augmented dataset. We noticed a tradeoff between precision and recall when we adjusted the confidence thresholds of bounding box output, with smaller thresholds resulting in more detections and larger thresholds filtering out some detections with lower confidence scores than the set thresholds. We suggested using smaller thresholds (approximately 0.5) for detection, counting, and measuring purposes to capture more “stomata” and “whole\textunderscore stomata”. The significant correlation between stomatal area, orientation, and the product of width and height and the ratio of stomatal width to height indicates the potential of using StoManager1 to measure leaf stomatal area and orientation. Our testing and validation of StoManager1's capabilities in detecting and measuring leaf stomatal properties for both the $Populus$ and hardwood datasets demonstrated its ability to accurately capture variation in stomatal properties across different hardwood species and genotypes within the same species. 

\subsection{Biological inference}\label{subsec2}

\begin{figure*}[!t]%
\centering
\includegraphics[width=\textwidth, angle=0]{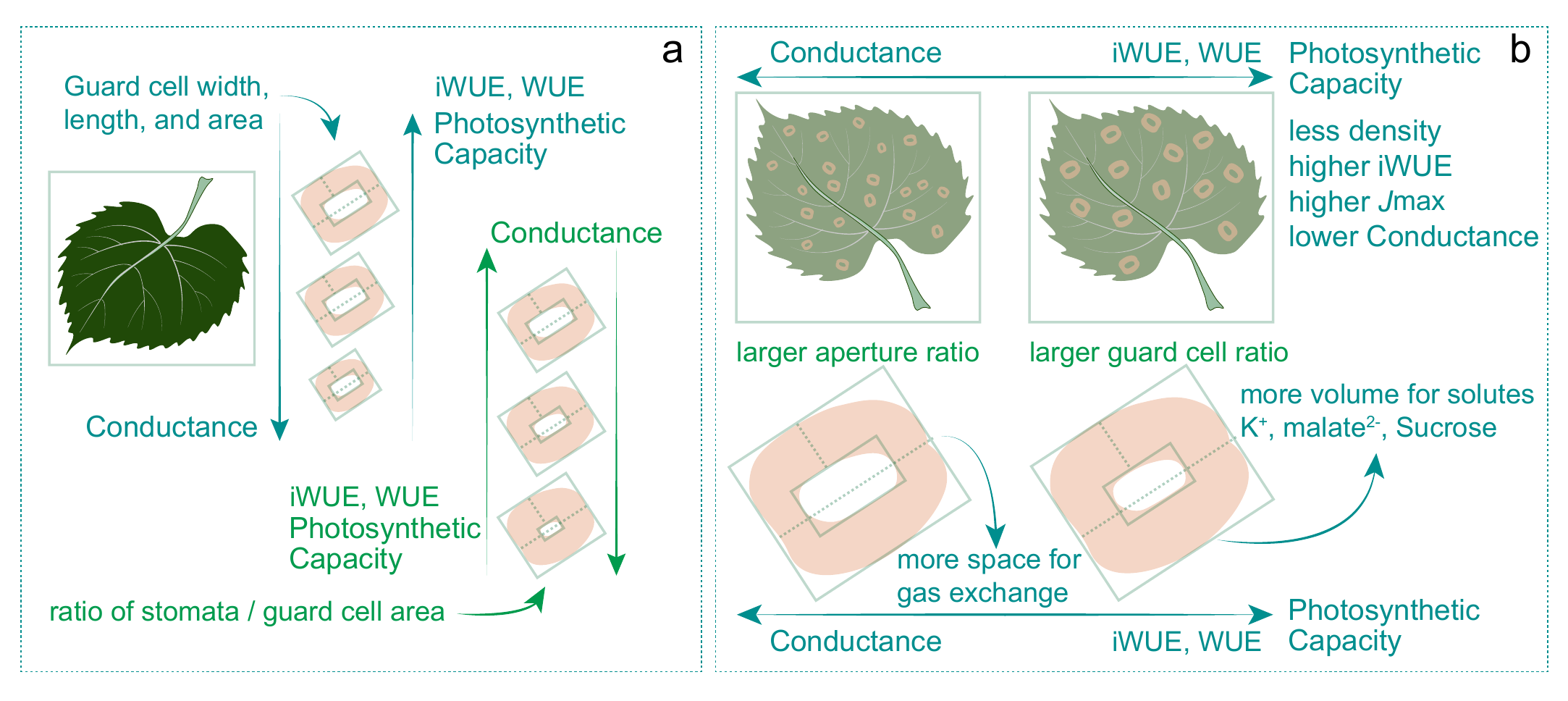}
\caption{Schematic diagrams of observed patterns using StoManager1-measured metrics and a) leaf functional traits, and b) potential mechanisms of how stomatal metrics regulate leaf physiology. WUE is instantaneous water use efficiency, and iWUE is intrinsic water use efficiency. Straight line arrows represent increasing trends.}\label{fig6}
\end{figure*}

Based on the results of $Populus$ dataset, we inferred that StoManager1-measured stomatal metrics play important roles in regulating leaf physiological and growth performance, and potential mechanisms are depicted in Fig. \ref{fig6}. Generally, our results suggested that StoManager1 measured stomatal metrics such as stomatal guard cell width, guard cell length, guard cell area, and the ratio of “stomata” area/guard cell area are significantly correlated with leaf $g_s$, iWUE, WUE, and photosynthetic capacity. Specifically, leaves with smaller stomata and guard cell area exhibited higher leaf $g_s$ but lower iWUE and $J_{max}$. The potential mechanism of this phenomenon would be that leaves with larger stomata and guard cells are more efficient in regulating stomatal opening and closure, and leaves with smaller stomata and guard cells tended to have a higher stomatal density and more open space for gas exchange to lose more water and take in more $\mathrm{CO}_2$ (Franks et al., 2015; Hughes et al., 2017). We also found that the ratio of “stomata” area/guard cell area exhibited contrasting effects on $g_s$ and iWUE and $J_{max}$. Specifically, leaves with larger ratio of “stomata” area/guard cell area had larger leaf $g_s$ and smaller iWUE and $J_{max}$, while leaves with smaller ratio of “stomata” area/guard cell area exhibited smaller leaf $g_s$ and larger iWUE and $J_{max}$. This may be attributed to the fact that higher “stomata” area ratio suggested more space or larger channels for gas exchange, while smaller ratio of “stomata” area/guard cell area indicates more guard cell volume for solutes such as $H^{+}$, $K^{+}$, $Cl^{-}$, $malate^{2-}$, and sucrose to more efficiently control stomata opening and closure regulating carbon and water cycling (Talbott \& Zeiger, 1998; Lawson \& Blatt, 2014).

Significant differences in StoManager1-measured metrics across different taxa of $Populus$ can be used as indicators or functional traits for identifying productive and water use efficient species for bioenergy production. The weighted multiple linear regression models suggested that StoManager1-measured metrics such as guard cell length significantly and positively contributed to $Populus$ productivity. As shown in PCA and Pearson correlation results, other StoManager1-measured metrics such as the ratio of “stomata area”/guard cell area, width, and length of “stomata” and “whole\textunderscore stomata”, and stomatal orientation relevant metrics could also potentially contribute to the leaf physiology and tree growth. 

Multiple linear regression models highlight the crucial role of stomatal metrics, measured by StoManager1, in regulating leaf physiology and tree growth in $Populus$. Employing weighted multiple linear regression models with leaf functional traits and leaf stomata metrics enabled us to account for 78\% and 93\% of the variances in productivity and intrinsic water use efficiency (iWUE), respectively. The findings suggest that stomatal metrics obtained through StoManager1 can be effectively utilized for modeling and estimating a plant's productivity and water use efficiency. This has significant implications for global gross primary productivity (GPP) modeling and estimation, as it proposes that integrating stomatal metrics into these models can enhance our comprehension and predictions of plant growth and resource usage on a worldwide scale. Furthermore, this research offers valuable insights for potential applications in forestry, agriculture, and ecosystem management. Gaining a better understanding of the factors affecting plant productivity and water use efficiency allows researchers to devise strategies for optimizing resource utilization, boosting crop yields, and encouraging sustainable management.

\subsection{Usage notes}\label{subsec2}

Using the StoManager1 model for leaf stomatal counting, measuring, and developing new indices can expedite stomatal and physiological research and increase sampling for more comprehensive studies. The model has vast potential in elucidating how leaf stomata regulate and control plant physiological performance and adapt to environmental stress and climate change. Using StoManager1 can offer valuable contributions to ecological modeling for global leaf functional traits, genetic modification of stomata for plants to cope with environmental stresses, and their role in climate change. Including leaf stomatal characteristics such as leaf stomatal and guard cell area, area variance, orientation, stomatal density, and leaf traits could lead to a more thorough understanding of the mechanisms underlying leaf function in regulating water and carbon cycling. StoManager1 provides an efficient tool for quick and accurate detection, counting and measuring stomatal metrics, thus increasing researcher productivity, and reducing data analysis time. Establishing empirical error rates can help assess measurement accuracy and improve the overall quality of research outcomes.

Future efforts to enhance StoManager1's capabilities include providing more training images from different species and families to enable researchers with large-scale and big datasets to conduct more comprehensive research. It should be noted that StoManager1, like any other tool or method, may have limitations or potential sources of error. To ensure accurate and reliable data, it is essential for users to thoroughly test the tool and compare its results with other methods, despite having tested and validated it on 17 hardwood species. Furthermore, it is crucial to consider any potential biases that could be introduced by the tool or how the data are collected and analyzed. StoManager1 shows excellent promise for researchers in ecology, plant biology and physiology, and other relevant fields. It has the potential to significantly improve the efficiency and accuracy of detecting, counting, and measuring stomata. We have developed Windows system-based standalone StoManager1 applications and made them freely available to the public on Zenodo (Wang et al., 2023a) and Figshare (Wang et al., 2023b).

Overall, we proposed that StoManager1, compiling bounding box- and segmentation-based models, has high precision, recall, and sensitivity and is well-suited for high-throughput stomatal characterization, specifically detection, counting, and measuring over 30 stomatal metrics. Using StoManager1, we developed new stomatal metrics, such as the ratio of “stomata area”/guard cell area, and these metrics showed significant correlations with leaf physiology and tree growth. StoManager1 allows researchers to investigate potential mechanisms of stomatal functioning. Using StoManager1, researchers can study plant functioning and adaptation on a large scale, providing valuable insights into how stomata regulate carbon and water cycling.

\section{Competing interests}
No competing interest is declared.

\section{Author contributions statement}

Jiaxin Wang: Conceptualization, Methodology, Formal analysis, Investigation, Data curation, Software, Validation, Visualization, Writing - original draft, Writing - review \& editing. Heidi Renninger: Project administration, Conceptualization, Funding acquisition, Supervision, Writing - review \& editing. Qin Ma: Writing - review \& editing. Shichao Jin: Writing - review \& editing.

\section{Acknowledgments}
This study was funded by the USDA National Institute of Food and Agriculture through the APPS grant (Advancing $Populus$ Pathways in the Southeast), with grant number 2018-68005-27636. The work was conducted as part of the Forest and Wildlife Research Center at Mississippi State University. The authors would like to acknowledge T. Hall, T. Durbin, A. Gentry, and H. Miles for their help with leaf collection, and Dr. Krishna Poudel at Mississippi State University for helping with data analysis and modeling. Additionally, Jiaxin Wang would like to express appreciation to Xuening Lu, Yansong Pei, Shangshang Wang, and Yehong Peng for their helpful coding assistance and company.

%\bibliographystyle{plain}
%\bibliography{reference}

%USE THE BELOW OPTIONS IN CASE YOU NEED AUTHOR YEAR FORMAT.
%\bibliographystyle{abbrvnat}
%\bibliography{reference}

\end{document}